\documentstyle[12pt,aaspp]{article}
\begin{document}

\def\pslandinsert#1{\epsffile{#1}}
\def\eqref#1{\ref{eq#1}}
\def\e#1{\label{eq#1}}
\def\be{\begin{equation}}
\def\ee{\end{equation}}
\def\ast{\mathchar"2203} \mathcode`*="002A
\def\rslash{\backslash} \def\oforder{\sim}
\def\larrow{\leftarrow} \def\rarrow{\rightarrow}
\def\darrow{\Longleftrightarrow}
\def\defeq{\equiv} \def\lteq{\leq} \def\gteq{\geq} \def\neq{\not=}
\def\<={\leq} \def\>={\geq} \def\lsls{\ll} \def\grgr{\gg}
\def\all{\forall} \def\lub{\sqcup} \def\relv{\vert}
\def\leftv{\left|} \def\rightv{\right|}
\def\%{\char'045{}}
\def\_{\vrule height 0.8pt depth 0pt width 1em}
\def\leftbrace{\left\{} \def\rightbrace{\right\}}
\def\sectsign{\S}
\def\prop{\propto}
\newbox\grsign \setbox\grsign=\hbox{$>$} \newdimen\grdimen \grdimen=\ht\grsign
\newbox\simlessbox \newbox\simgreatbox
\setbox\simgreatbox=\hbox{\raise.5ex\hbox{$>$}\llap
     {\lower.5ex\hbox{$\sim$}}}\ht1=\grdimen\dp1=0pt
\setbox\simlessbox=\hbox{\raise.5ex\hbox{$<$}\llap
     {\lower.5ex\hbox{$\sim$}}}\ht2=\grdimen\dp2=0pt
\def\simgreat{\mathrel{\copy\simgreatbox}}
\def\simless{\mathrel{\copy\simlessbox}}
\def\spose{\rlap} \def\caret#1{\widehat #1}
\def\hat#1{\widehat #1} \def\tilde#1{\widetilde #1}
\def\limitswitch{\limits} \def\dispstyle{\displaystyle}
\def\dot#1{\vbox{\baselineskip=-1pt\lineskip=1pt
     \halign{\hfil ##\hfil\cr.\cr $#1$\cr}}}
\def\ddot#1{\vbox{\baselineskip=-1pt\lineskip=1pt
     \halign{\hfil##\hfil\cr..\cr $#1$\cr}}}
\def\dddot#1{\vbox{\baselineskip=-1pt\lineskip=1pt
     \halign{\hfil##\hfil\cr...\cr $#1$\cr}}}
\def\Abf{{\bf A}}\def\ybf{{\bf y}}\def\Ebf{{\bf E}}\def\cbf{{\bf c}}
\def\Zbf{{\bf Z}}\def\Bbf{{\bf B}}\def\Cbf{{\bf C}}\def\bfB{{\bf B}}
\def\bfq{{\bf q}}\def\bfc{{\bf c}}\def\bfy{{\bf y}}\def\bfv{{\bf v}}
\def\bfp{{\bf p}}

\def \GRBs{$\gamma$-ray bursts }
\def \GRBsb{$\gamma$-ray bursts}
\def \GRB{$\gamma$-ray burst }
\def \NSs{neutron stars }
\def \NS{neutron stars }
\def \Paczy{Paczy{\'n}ski }
\def \paczy{Paczy{\'n}ski }
\def \etal{{\it et al. }}
\def \MC{Magellanic Cloud }
\def \MCs{Magellanic Clouds }
\def \MCsb{Magellanic Clouds}
\def \MStream{Magellanic Stream }
\def \MStreamb{Magellanic Stream}
\def \Mstreamb{Magellanic Stream}
\def \Mstream{Magellanic Stream }
\def \MGroup{Magellanic Group }
\def \MGroupb{Magellanic Group}
\def \Mgroup{Magellanic Group }
\def \Mplane{Magellanic plane }
\def \MPlane{Magellanic plane }
\def \MW{Milky Way }
\def \LL81{Lin and Lynden-Bell 1981}
\def \MF80{Murai and Fujimoto 1980}
\def \vovmax{$\left<V/V_{max}\right>$}
\def \vovmaxb{$\left<V/V_{max}\right>\,\,$}
\def \cmaxcmin{C_{max}/C_{min}}
\def \degrees{^{\circ}}

\title{Indications for $\gamma$-ray Bursts Originating Within \\
an Extended Galactic Halo ?}

\vspace{0.3in}

\author{Eyal Maoz}
\vspace{0.3in}

\affil{Harvard-Smithsonian Center for Astrophysics, \\
MS 51, 60 Garden Street, Cambridge, MA~02138}

\vspace{0.2in}
\centerline{E-mail: maoz@cfa.harvard.edu}
\vspace{0.6in}
\centerline{Submitted to the Astrophysical Journal Letters}
\vspace{0.8in}

\begin{abstract}
If a substantial fraction of the observed \GRBs
originates within an extended Galactic
halo then their spatial distribution should deviate slightly from
spherical symmetry in a very particular way which involves
features both in the bursts' angular and radial distributions.
This conclusion is based on various reasons
which are all related to the presence and motion of the satellite galaxies
around the Galaxy, and is
independent of the nature and origin of the sources.

We analyze the spatial distribution of the bursts, according to
the BATSE catalog until March 1992, and argue that the expected
signature of an extended Galactic halo model is indicated by the data.
The distance to the faintest bursts in the halo is either $\sim130$\,Kpc
or $\sim270$\,Kpc.

Although a signature of a nearby-extragalactic distance scale in the data
is very suggestive, we argue that a comparison with specific models
is necessary before regarding our findings as a conclusive evidence.
If the increasing data supports our results then \GRBs may
be the first detected manifestations of nearby intergalactic objects, either
primordial or which have escaped predominantely from our satellite galaxies.

\end{abstract}
\newpage

\section{INTRODUCTION}
Two decades after the discovery of \GRBs (Klebesadel, Strong, and Olson
1973) the origin of these events is still an enigma.
There are some indications
that \GRBs may originate on or near neutron stars (Mazets 1988; Murakami \etal
1988; Fenimore \etal 1988) but there is no consensus on a
distance scale, not even to within orders of magnitude (\paczy 1992).
According to the observations made with the BATSE experiment on the {\it
Compton Gamma-Ray Observatory (GRO)}, \GRBs are distributed isotropically over
the sky (Fishman \etal 1991; Meegan \etal 1992)
and do not show any association with known astronomical populations
such as a concentration towards the Galactic plane, a correlation with the
Galactic center, the \MCsb, or with prominent extragalactic regions such as the
Virgo cluster.
A single Galactic disk distribution seems to be ruled out from the
angular distribution of weak \GRBs (Mao and \paczy 1992a), and
the Oort cloud of comets is also an unlikely source of the bursts (Maoz 1993).
The observed bursts' distribution is consistent with a cosmological
distance scale (e.g. Fenimore \etal 1992; \paczy 1992; Mao and \paczy 1992b;
Piran 1992; Dermer 1992; \Paczy 1991;
Kouveliotou \etal 1992; Paciesas \etal 1992; Norris \etal 1992),
or with an extended Galactic halo distribution (see detailed discussion and
references in \S{2.2}).

We argue that if the observed \GRBsb, or at least
a substantial fraction of them, originate within an extended Galactic
halo then their spatial distribution {\it should\/} slightly deviate from
spherical symmetry in a very particular way, regardless of the nature
and origin of the bursting objects. This is based on a variety of reasons
which are all related to the presence and motion of the satellite galaxies
around our galaxy (\S{2}). In \S{3} we show that
a detection of the signature for a nearby-extragalactic distance scale
is very suggestive. We then discuss the possibly emerging picture
(\S{4}) and make predictions (\S{5}).

\section{IMPRINTS IN A HALO DISTRIBUTION}
\subsection{The Magellanic Planes}
First, let us discuss the existence of two special planes which will turn
out to be relevant to \GRBs in \S{2.2}.

The \MStreamb, a narrow band of neutral hydrogen gas (e.g. Mihalas and Binney
1981) defines a great circle on the sky. Various analyses
(\MF80; \LL81 and references therein) have led to the
following, generally accepted, picture:  The
Stream consists of material torn out from the \MCs by the Galaxy's tidal
field during the previous close passage, and lies in their orbital plane.
The \MCs have been orbiting
together in this plane as a binary system for a long time, their orbit
is eccentric and
runs roughly between $50\hbox{-}120\,$Kpc from the
galactic center at the present epoch.
The normal to this plane (hereafter, the {\it \MStream plane\/}, or the
{\it MS-plane\/}) points to the direction
$(l,b)\hbox{=}(185\degrees,3\degrees)$ (\LL81), where $l$ and $b$
are the Galactic longitude and latitude, respectively.
Thus, the MS-plane
is almost perpendicular to the Galactic plane and is orthogonal to the line
joining the Sun to the Galactic center.

Our galaxy has several other companions which could in principle
be orbiting each in a different plane. However,
Kunkel and Demers (1976) noticed that most of these
dwarf spheroidal galaxies and distant
globular clusters appear to lie very close to a great circle on the sky
(see also Lynden-Bell 1976a,b,1982a; Fich and Tremaine 1991).
This includes Leo I, Leo II, Draco, Ursa Minor, NGC 7006, Pal 3,4, and 12
(hereafter, the {\it \MGroupb}).
This plane is likely to be the common orbital plane of these satellites
since, on one hand,
the probability that such a planar distribution might arise by
chance is less than $0.002$ (Kunkel and Demers
1976), and on the other hand, there are a few coincidences which indicate some
physical association between these satellites such as
structural elongation along the circle (e.g., Lynden-Bell 1982a), and
longitudal distribution in this plane which matches the expected from dynamical
considerations (e.g., Hunter and Tremaine 1977).
This roughly planar
distribution may be explained by a a breakup of a larger satellite during
a past tidal interaction with the Galaxy
which led to the strewn distribution of smaller systems over its orbital plane
(Toomre 1974; Lynden-Bell 1982b).
The normal to this plane (hereafter the {\it \MGroup plane\/} or the
{\it MG-plane\/}) points to the direction
$(l,b)\hbox{=}(169\degrees,-23\degrees)$ (Kunkel and Demers 1976).
It differs from the normal to the MS-plane
by $40\degrees$, but the \MCs lie only $7\degrees$ from it (Figure 1).

\subsection{The Expected Imprints}
There are suggestions for \GRBs originating in an extended Galactic halo of
\NSs (e.g.,
Fishman \etal 1978; Jennings and White 1980; Shklovski and Mitrofanov 1985;
Atteia and Hurley 1986; Jennings 1984; Yamagami and Nishimura 1986).
These \NSs could be born in the Galactic disk and ejected at high velocities,
either due to asymmetric explosions or due
to the unbinding of a binary system during a supernova explosion, and in this
way populate a large spherical region (Lyne, Anderson, and Salter 1982; Cordes
1986 ; Hartmann, Epstein, and Woosley 1990), but
there are some difficulties with this idea (Mao and \paczy 1992a;
\Paczy 1991).
In general, if the conceivable bursting objects have migrated away from the
Galactic disk forming an extended halo we may expect also
similar objects escaping from the \MCs and from the other satellite galaxies.
These satellites may contribute to the hypothesized
extended halo more than the expected from their masses (relative to the disk's
mass) since their escape velocities are lower.
Fabian and Podsiadlowski (1993) have even suggested that {\it most\/}
of the bursting objects originate in the \MCsb.
The important point is that the spatial distribution of these objects
will {\it not\/} be spherically symmetric, but should show some
concentration towards the orbital plane of the satellite they have escaped
from.  We do not expect this enhancement to look like a narrow disk but it
should have some oblate shape with the principle
plane coinciding with the satellite's orbital plane.

This general idea
does not necessarily require having an efficient ejection
mechanism for the objects. Tidal fields from the Galaxy
during a satellite's close
passage (Lynden-Bell 1976a; Fujimoto and Sofue 1976; \MF80), or interactions
between the satellites themselves would lead to
a strewn distribution of matter (and also of
potentially bursting objects) over the orbital planes in a similar way to the
recent formation of the \Mstreamb. Again, the detached material
need not necessarily form a very narrow ring since, for example,
the spin
axis of the LMC lies close to the orbital plane, so angular momentum of the LMC
can carry detached material some distance out of the MS-plane.

It is also possible that \GRBs are associated with an extended Galactic halo
of primordial objects (e.g. Eichler and Silk 1992) which could be
isotropically distributed. However,
regardless of their origin, the spatial distribution of any conceivable
objects
must be gravitationally distorted due to their interaction with the satellite
galaxies which are traveling through this halo.
Each satellite induces a density perturbation in the halo which can be viewed
as the combination of tidal forces (dynamical tides) and a wake of density
enhancement trailing behind the moving satellite (which generates dynamical
friction). The response of the halo distribution to the gravitational
perturbation of an orbiting satellite has a complex pattern but it always
involves some concentration of halo particles towards the satellite's
orbit (e.g., Weinberg 1989).

All the above arguments essentially
predict some concentration of bursting objects towards
the satellites' orbits. This implies an enhancement in their angular
distribution towards the MS-plane and the MG-plane, but also some overdensity
in their radial distribution at the distances at which
the satellites are orbiting. We shall now define the signature that
should identify an extended Galactic halo model.

\subsection{The Signature of an Extended Halo}
Testing an extended Galactic halo model for the distribution of \GRBs
is not straightforward as,
for example, the $\left<\sin^{2}b\right>$ statistics applied for testing a
Galactic disk model, or the dipole test, $\left<\cos\theta\right>$,
for a (non-extended) Galactic halo model.
In our case,
the expected magnitude and pattern of the deviations from spherical symmetry
should be computed for specific models (either analytically or using
simulations) and confronted with observations. Such model-dependent tests
would probably involve some free parameters and require
more data than currently available.
However, all the various reasons for expecting deviations from
spherical symmetry (\S{2.2}) predict very similar patterns of
distortion in the bursts' distribution, and
these common features define
the {\it first-order signature\/} of an extended Galactic halo model.

The expected signature in the bursts' distribution is thus the following:
a) There should be some concentration of bursts towards the two
planes (\S{2.1}), but since the \MGroup satellites seem to orbit at larger
distances than the \MCs (e.g. Binney and Tremaine 1987)
the imprints of the MG-plane should be dominant
at higher distances, and that of the MS-plane at smaller distances.
Therefore, if the observed \GRBs originate
within $\lesssim{80}\,$Kpc then we expect to detect
some concentration in their angular distribution mainly towards
the MS-plane, but if their spatial distribution extends
further out then a concentration should show up towards the MS-plane
only within small relative distances, but also towards the MG-plane over
larger relative distances.
b) Regarding the radial distribution, we expect some concentration of bursts
within the range of distances inside which the dominant satellites are
orbiting (\S{2.2}). This does not mean an increase in the burst number
density at the corresponding
radii, but that the density profile should become less
steep at those distances.

\section{COMPARISON WITH OBSERVATIONS}
We use the recently published BATSE catalog at the GRO Science
Support Center which includes the locations and count rates of 241 bursts
observed until March 1992 (Fig. 1).  Assuming "standard candles" (see
discussion in \S{5}) we
construct a data set of the 3-D bursts' locations sorted by
relative distance, $D_{i}\propto(\cmaxcmin)_{i}^{-1/2}$,
where $0<D_{i}\leq 1$, $i=1,..,241$.

Figure 2 shows the concentration towards both planes of \GRBs
which originated {\it within\/} an increasing relative distance.
Apparently, there is a concentration of bursts
towards the MS-plane, but only of close (strong) ones (Fig.2-a). However,
the same plot with respect to the MG-plane (Fig.2-b) shows that
the entire
curve rises considerably above the expectation value, indicating a
concentration of bursts towards the MG-plane essentially at all radii.
This is precisely the signature of
a {\it very\/} extended Galactic halo distribution (i.e. a nearby-extragalactic
origin) of \GRBs (\S{2.3}),
namely, a concentration only of relatively close bursts towards the
MS-plane, along with a concentration towards the MG-plane over larger
radial distances.

Although these plots are cumulative ones, their shapes
still reflect variations in the degree of concentration towards the planes
due to the rapidly increasing
density of data points on the curves with distance. In order to demonstrate
that the cumulative nature of these plots does not introduce a severe
artifact, we grouped the bursts into five independent
bins according to their radial distance and show that the
results nicely agree with the expected signature
in the angular distribution (Table 1).

Regarding the radial distribution, Figure 3 shows the \vovmaxb parameter as
a function of sample depth (see also Atteia and Dezalay 1993) and thus
gives an estimate for {\it how fast\/} the bursts spatial density falls
with distance. For example, a value of 0.5 indicates a constant density, 0.4
corresponds to $\sim{r^{-1}}$ falloff, and 1/3 to $\sim{r^{-3/2}}$.
Apparently, the bursts' density profile varies
from roughly constant at small distances to roughly $\sim{r^{-3/2}}$ at
large distances, but the logarithmic slope does {\it not\/} seem to change
monotonically. We notice that there are interesting coincidences between
the distances at which \vovmaxb increases and the distances at which
the concentration towards the two
planes is relatively high: both
curves (Fig. 2 and 3) rise between $0.4\le D\le 0.45$, both of them
(Fig. 2-a and 3)
show concentration higher than the expected from a monotonic decline between
$0.4\lesssim D\lesssim 0.6$, and similarly at $0.15\lesssim D\lesssim 0.22$
in which the density profile even increases with distance (\vovmax $>0.5$).
These coincidences could arise by chance, but the fact is that they are
expected on theoretical grounds in any extended Galactic halo scenario
(\S{2.3}).

We have {\it qualitatively\/} identified in the data the first order
signature (\S{2.3}) of an extended Galactic halo model, but
a rigorous evaluation of the statistical significance of this signature
requires confronting our results with quantitative predictions of
specific models.
Since detailed models are out of the scope of this paper, we shall just
draw the attention to the following points: a) the curve in Figure 2-b is
{\it consistently\/} above the expectation value (see also Table 1) which
implies a concentration of bursts towards the MG-plane over a
{\it wide} range of distances, as indeed expected. The statistical
significance for such
concentration depends on the sample depth and varies roughly
between $1\hbox{-}2\sigma$ (Fig. 2-b). b) The distribution of 160 bursts
of pre-BATSE data obtained from the KONUS experiment (Mazets, \etal 1981)
also show a concentration ($1.4\sigma$) towards the MG-plane (Table
1), but it is unclear whether the sky exposure for this data is uniform
enough for taking this seriously.
c) There {\it is\/} a concentration of bursts also towards the MS-plane, but
{\it only\/} of relatively close ones, exactly as expected.
d) There seem to be correlations beween features in the radial and the angular
distributions, as indeed expected if they are due to the presence and motion
of the satellite galaxies.

These findings {\it cannot\/} be taken as a conclusive evidence at this stage,
but a detection of the signature for a nearby-extragalactic distance scale in
the bursts' distribution is {\it definitely suggestive}.

\section{THE SUGGESTED PICTURE}
Assuming that the increasing data supports our findings, we may identify
the distance to the \MCs ($\sim 60$\,Kpc) either
with the peak (Fig.2) at $D\sim 0.45$ or with the one at $D\sim 0.22$
(the \MCs are the major satellite, they lie
very close to both planes,
and we are not aware of any other closer substantial structures).
Normalizing the distance scale accordingly we find that
the current maximum sampling depth ($D\hbox{=}1$) is either $\sim130\,Kpc$ or
$\sim270\,Kpc$, respectively, but
this is only a crude estimate since the clouds' orbit is eccentric (Lin and
Lynden-Bell 1981). b) Assuming that most bursts originate in this extended
halo we find that their
rate is $\sim10^{-5}\,yr^{-1}(Kpc)^{-3}$ in our near vicinity,
so we do not expect an enhancement of bursts from the LMC itself, neither
from any visible part of other galaxies (see discussion in \S{5}).

Our findings support the picture that a fraction of the bursting objects
have escaped (or have been detached) from our satellite galaxies, but
it is also consistent with the idea of a primordial population of bursting
objects.
We find the latter possibility especially interesting
for the following reason:
if the observed roughly planar distribution of our satellite galaxies
(the MG-plane) is due to some primordial conditions in the surrounding mass
density field, then it is reasonable to expect also some concentration of the
hypothesized primordial objects towards this plane.
Furthermore, simulations of halo formations (Dubinski and Carlberg 1991)
show strong tendency towards triaxial shapes, with the minor
axis in the direction of the angular momentum vector (Dubinski 1992).
This nicely coincides with the observed indication for a slightly
nonspherical distribution with a minor axis normal to the MG-plane, i.e.,
parallel to the angular momentum vector of the satellites' orbit.
It is also interesting to notice that
M31's {\it initial\/} direction, i.e.,
the  direction ($l\hbox{=}101\degrees,b\hbox{=}-27\degrees$)
from which M31 had emerged during its formation epoch
(Lynden-Bell and Roychaudhury 1989) is very close to the MG-plane.
This strengthens our suggestion that this plane is of special physical
importance
throughout the entire Local Group, and thus should also show up in the
bursting objects' distribution if they are primordial or
left over from the formation epoch of the Local Group of galaxies. Thus,
\GRBs may well be associated with intergalactic objects such
as primordial
black holes or old dense stellar systems (e.g., Eichler and Silk 1992).

\section{DISCUSSION AND PREDICTIONS}
The expected signature (\S{2.3}) for an extended halo of sources implies that
applying statistical tests only for the
entire distribution of the observed bursts
as a whole is inadequate. Furthermore,
if some of the observed bursts originate at distances which
are beyond the scale over which the MG-plane is expected to make an
imprint, this would smear out the expected anisotropic features.
Therefore, we had to examine the degree of burst concentration
towards both planes as a function of increasing sample depth.

The assumption  of ``standard candles'' need not
necessarily be adequate, but if the peaks in
Figure 2-b will not get much broader with the increasing data then it would
indicate that the luminosity function is probably
quite peaked for the following reason: wide-range luminosity
functions (e.g., shallow power-laws) would smear out any real structure
in the radial distribution. Afterall,
the variable $D\equiv(\cmaxcmin)^{-1/2}$ is actually a convolution of the real
distance to the sources with the luminosity function, and
a convolution with a slowly changing function erases high frequency features.

We might be on the edge of observing the halo of bursting objects around
M31, but this strongly
depends on how peaked is the bursts' luminosity function.
We predict that a concentration of bursts will {\it rapidly\/} grow in that
direction if the bursts' detection limit is reduced by $\sim1.5$
orders of magnitude (at the expanse of lowering the statistical significance
of identifying a real burst).
We also predict that the conventionally applied  $\left<\sin^{2}b\right>$
test in {\it Galactic\/} coordinates will keep
indicating a weak concentration of all bursts towards the Galactic
disk with low
significance level due to a simple reason: we see in Figure 1 that the MG-plane
runs only between the Galactic latitudes $\sim\pm70\degrees$.
Thus, if bursts are concentrated towards the MG-plane they must
be, on the average, also slightly closer to the Galactic disk than the
expected from an isotropic distribution.

Finally, we should bear in mind that the data is also statistically
consistent with a featureless isotropic distribution. However, the fact is
that the (small) anisotropies in the angular distribution and the variations
in the radial distribution agree very well with what is theoretically
expected if \GRBs (or at least a substantial fraction of them) originate in
an extended Galactic halo.
If the detected signature (\S{2.3}) will not disapear with the increasing
data then it would strongly support our suggested distance scale.
The possibility
that the distribution of bursting objects is similar to that of the
dark matter, and thus might be related to it
in some way, is especially exciting.

I wish to thank William Press, George Field, and John Dubinski for
discussions and comments, and Alar Toomre for an enlightening discussion.
This work was supported by the U.S. National Science Foundation, grant
PHY-91-06678.

\newpage
\vspace{2.3in}

{\bf Fig. 1} - An equal-area (Aitoff) projection of the 241
bursts' locations from the BATSE catalog on the entire celestial sphere in
galactic coordinates.
The orbit of the \MCs (empty triangles) runs very close to the $90\degrees$ and
$270\degrees$ longitude lines (the MS-plane). The clouds and
the other members of the \MGroup (small empty circles) define the MG-plane
(the dotted curve).  The location of the Andromeda
galaxy (M31) is also shown as a larger circle near the lower left corner.

\vspace{0.2in}

{\bf Fig. 2}
(a) Each point on the curve
indicates the degree of concentration towards the MS-plane for bursts which
originated {\it within\/} a relative distance $D_n$,
where $\left<\cos^{2}b'\right>_{n}\equiv n^{-1}\sum_{i=1}^{n}
{\cos^{2}(b_{i}')}$, $n\hbox{=}3,..,241$,
and $b'$ is the the angular distance to the plane.
The higher this value is above $2/3$, the stronger is the concentration.
The error bars reflect one standard deviation from the
expectation value for an isotropic distribution, and are
given by $(45n/4)^{-1/2}$.
(b) - A similar plot relative to the MG-plane. There is
some concentration of bursts towards the MG-plane over large
relative distances, and towards the MS-plane only at small distance, exactly
as expected (\S{3}).
Surprisingly, taking the non-uniform sky coverage into account (a function
of declination) results only in marginal changes in both plots: the
curves exactly maintain their shapes but are lowered by $\sim0.1\sigma$ on the
average, and nowhere change by more than $0.24\sigma$.

\vspace{0.2in}

{\bf Fig. 3}
The \vovmaxb parameter as a function of sample depth. There is some correlation
between the distances at which \vovmaxb increases and the distances at which
the concentration towards both planes is higher than the average, as indeed
expected in an extended halo model (see \S{3}).

\newpage
\vspace{1.5in}

{\bf TABLE 1} - Concentration of Independent Samples towards the two planes

\vspace{0.5in}

\begin{tabular}{l r c c c}
 relative &number of & & &one standard\\
 distance &bursts\,\,\,\,\,\,\, &$\left<\cos^{2}b_{MG}'\right>$ &
 $\left<\cos^{2}b_{MS}'\right>$ &deviation \\ \hline\hline
 0.0-0.2 &12  &0.738  &0.747 &0.086\\
 0.2-0.4 &37  &0.720  &0.645 &0.049\\
 0.4-0.6 &71  &0.676  &0.670 &0.035\\
 0.6-0.8 &53  &0.704  &0.637 &0.041\\
 0.8-1.0 &67  &0.688  &0.609 &0.036\\ \hline
 KONUS   &160 &0.699  &0.674 &0.023\\
\end{tabular}

\vspace{1.0in}

{\bf Table 1} -
The BATSE data binned into five independent sets according to ranges of
distance,
and the sample of $160$ bursts obtained from the KONUS (Mazets \etal 1981)
experiment aboard the Soviet satellites Venera 11-14.
All the sets show various degrees of concentration
towards the MG-plane ($\left<\cos^{2}b'\right>$ higher than $2/3$)
but only the close ones show some concentration towards the MS-plane,
exactly as expected (see \S{2.3}).

\end{document}